\documentclass[aps,pra,preprint,groupedaddress,showpacs]{revtex4}
\usepackage{graphicx} 
\usepackage{dcolumn}  
\usepackage{bm}       
\begin{document}
\preprint{Version 3.0}

\title{Radiative Corrections to Parity Nonconserving \\ Transitions in Atoms}

\author{J. Sapirstein}
\email[]{jsapirst@nd.edu}
\affiliation{
Department of Physics, University of Notre Dame, Notre Dame, IN 46556}

\author{K. Pachucki}
\email[]{krp@fuw.edu.pl}

\author{A. Veitia}
\email[]{aveitia@fuw.edu.pl}
\affiliation{
Institute of Theoretical Physics, Warsaw University, Hoza 69, 00-681,
Warsaw, Poland}

\author{K. T. Cheng}
\email[]{ktcheng@llnl.gov}
\affiliation{
University of California, Lawrence Livermore National Laboratory,
Livermore, CA 94550}

\date{\today}

\begin{abstract}
The matrix element of a bound electron interacting with the nucleus
through exchange of a Z boson is studied for the gauge invariant case
of $2s_{1/2}-2p_{1/2}$ transitions in hydrogenic ions. The QED radiative
correction to the matrix element, which is $-\alpha/2\pi$ in lowest order,
is calculated to all orders in $Z\alpha$ using exact propagators. Previous
calculations of the first-order binding correction are confirmed both
analytically and by fitting the exact function at low $Z$. Consequences
for the interpretation of parity nonconservation in cesium are discussed.
\end{abstract}


\maketitle

\section{Introduction}

The calculation of radiative corrections in atoms with low nuclear
charge $Z$ is facilitated by the fact that binding corrections, which
enter as powers and logarithms of $Z \alpha$, are relatively small,
and can be treated in perturbation theory. For atoms with high nuclear
charge the perturbation expansion converges more slowly, and for
highly-charged ions the expansion is generally avoided, which is possible
when numerical methods are used to represent the electron propagator.
This approach, first introduced by Wichmann and Kroll \cite{W-K}
for the vacuum polarization and Brown and Mayers \cite{Brown} for the
self-energy, has been applied to the calculation of both energy levels,
notably by Mohr and collaborators \cite{PJM}, and more recently to matrix
elements, specifically hyperfine splitting (hfs) and the Zeeman effect
\cite{hfsallZ,BCS}.

It is of interest to further extend this kind of radiative
correction calculation to the parity nonconserving (PNC) process
$6s_{1/2} \rightarrow 7s_{1/2}$ in neutral cesium \cite{Wieman}.
Corrections to this transition are of importance for the question of
whether a breakdown of the standard model is present for cesium PNC.
Specifically, if the radiative correction to the electron-Z vertex is
taken to be its lowest-order value, $-\alpha / 2 \pi$, then based on
the the present status of other corrections to PNC which have included
a number of significant shifts only recently considered that arise from
the Breit interaction \cite{Breit} and vacuum polarization \cite{vacpol},
a discrepancy with experiment of approximately $2 \sigma$ would result.
Given the presence of other indications of possible problems with
electroweak tests of the standard model, specifically the NuTev result
\cite{NuTev} and hadronic asymmetries in $Z \rightarrow b \bar{b}$
\cite{hadasy}, a discrepancy in cesium PNC could be an indication of
new physics.

However, it is known that binding corrections to the similar matrix
element involved in hfs are very large for highly-charged ions.
That this is so is not surprising, given the first two terms of the
one-loop vertex correction to hfs \cite{BCS},
\begin{equation}
\delta \nu = {\alpha \over \pi} E_F \biggl[ {1 \over 2} +
\biggl( {\rm ln} 2 - {13 \over 4}\biggr) \pi Z \alpha\biggr],
\label{eq:eq1}
\end{equation}
where $E_F$ is the lowest-order hfs energy.
Already at $Z=9$ the leading binding correction leads to a change
in sign of the hfs, and at $Z=55$ the formula would predict
$-2.72 \,{\alpha \over \pi}\, E_F$, as compared to the low-order,
uncorrected value of  $+0.5{\alpha \over \pi}\, E_F$. Of course,
with $Z \alpha$ = 0.4, the above equation, even with known higher-order
terms included, cannot replace an exact evaluation. As mentioned above,
such evaluations have been carried out by a number of groups, and the
complete answer turns out to be $-3.02\, {\alpha \over \pi}\, E_F$
\cite{BCS}.

It is possible to carry out a parallel analysis for radiative corrections
to PNC. If we define the lowest-order PNC matrix element as $Q_0$ and the
one-loop radiatively corrected matrix element as $Q_R$, with
\begin{equation}
Q_R ={\alpha \over \pi}\,  Q_0\, R(Z \alpha),
\end{equation}
the first two terms of $R(Z\alpha)$ are
\begin{equation}
R(Z\alpha) =  -{1 \over 2} - \biggl( 2\, {\rm ln} 2 + {7 \over 12}\biggr)
\pi Z \alpha,
\label{eq:perturb}
\end{equation}
where the first term is part of the standard radiative correction for
atomic PNC \cite{Marciano-Sirlin} and the leading binding correction was
first calculated in Ref.~\cite{Milstein}. For the case of cesium this
formula changes the coefficient of $\alpha / \pi$ from -0.5 to -2.98,
changing a negligible -0.12 percent to a significant -0.69 percent shift.
This largely removes the $2 \sigma$ discrepancy between theory and
experiment.

There are a number of issues that must be addressed before accepting the
-0.69 percent shift at face value.  Firstly, just as with hfs, an approach
that does not rely on expansion in $Z \alpha$ is required. Even though the
first two terms in Eq.~(\ref{eq:eq1}) for the vertex correction to hfs give
an answer within 12 percent of the total answer, there is no reason we know
of for this to be true in general. Secondly, it is not clear that it is
correct to use $Z=55$ in the above equation. When the cesium 6s Lamb shift,
which is also governed by short distance effects, is studied with all-orders
methods \cite{Pyykko,CS}, a much smaller effective nuclear charge is seen,
specifically about 14. Thirdly, an important difference between PNC and hfs
is the role of gauge invariance. In the latter case the initial and final
states are real physical states.  However, the Z boson vertex does not
involve two physical states, instead involving either a $6s_{1/2}$ or
$7s_{1/2}$ state and an intermediate state with $p_{1/2}$ quantum numbers.
While it can be shown that Eq.~(\ref{eq:perturb}) is still valid in this
case, higher-order binding corrections will be gauge dependent.

To address the last issue, we choose here to work with a gauge-invariant
quantity, the matrix element of the weak Hamiltonian
\begin{equation}
H_W = Q_W {G_F \over \sqrt{8}} \gamma_0 \gamma_5 \rho_{N}(\vec r)
\end{equation}
between the $2s_{1/2}$ and $2p_{1/2}$ states of a hydrogenic ion, where
$\rho_N(\vec r)$ describes the distribution of the weak nuclear charge,
which is close to the neutron distribution. While a finite distribution
will be used for $\rho_N(\vec r)$, the atomic $2s_{1/2}$ and $2p_{1/2}$
states will be chosen to be solutions of the Dirac equation with a point
nucleus, so the energies of these two states are equal. This allows
radiative corrections to PNC to be studied nonperturbatively to all orders
in $Z\alpha$ in a manner parallel to that used for hfs \cite{BCS,NewCS},
and in particular gives information about the $Z\alpha$ behavior of
the function $R(Z\alpha)$ that will be useful when the cesium problem
is addressed, as will be discussed in the conclusion.

The plan of the paper is the following. The lowest-order matrix element
$Q_0$ is treated in Sec.~\ref{sec:lowest}. In Sec.~\ref{sec:derivation}
we give a derivation of the radiative correction formulas, and in
Sec.~\ref{expansion} evaluate $R(Z\alpha)$ to first order in $Z\alpha$,
confirming the result of Ref.~\cite{Milstein}. In Sec.~\ref{sec:numerical}
we rearrange the formulas in a way that allows for an exact numerical
evaluation, and present the details of such a calculation for the range
$Z=10-100$. In the last section, it is shown that the numerical evaluation
at low $Z$ agrees with the perturbative expansion, and the higher-order
binding corrections inferred. Prospects for extension of the calculation
to the actual experiment, where a laser photon is present driving the
$6s_{1/2}-7s_{1/2}$ transition, are also discussed.

\section{\label{sec:lowest}Lowest-order calculation}

The matrix element of the weak charge operator in lowest order is
\begin{equation}
Q_0 \equiv Q_{wv} = \int d^3 r \psi_w^{\dagger}(\vec r) \gamma_5
\psi_v(\vec r) \rho_N(\vec r),
\end{equation}
where we shall from now on suppress the overall factor $Q_W G_F / \sqrt{8}$,
use $w$ to denote the $2p_{1/2}$ state, and $v$ the $2s_{1/2}$ state.
The nuclear distribution is chosen to be uniform, with a radius $R_0$ fixed
so that the root-mean-square radius agrees with a fermi distribution with a
thickness parameter 2.3 fm and a $c$ parameter given in Table \ref{tab:tab1}.
Because of the simplicity of the uniform distribution the matrix element
can be evaluated analytically, and is
\begin{equation}
Q_0 = {6 i Z^4 \alpha \over \pi N_2^5} \biggl({2 Z R_0 \over N_2 a_0}
\biggr)^{\!2 \gamma -2} \!{\sqrt{1+2\gamma} \over \Gamma(2\gamma+2)} \,
e^{-{2ZR_0 \over N_2 a_0}} a_0^{-3}.
\end{equation}
Here $\gamma = \sqrt{1-(Z\alpha)^2}$, $N_2 = \sqrt{2(1+\gamma)}$, and
$a_0$ is the Bohr radius. We note the singularity of this expression as
$R_0 \rightarrow 0$, which at small $Z$ manifests itself as a logarithmic
dependence on $R_0$, as can be seen from the Taylor expansion in
$Z\alpha$ of the above,
\begin{equation}
Q_0 = {\sqrt{3}i Z^4 \alpha \over 32 \pi} e^{-x}
\biggl[ 1 + (Z\alpha)^2 \biggl( -{\rm ln} x - \gamma_E +
{55 \over 24} - {x \over 8}\biggr) + O (Z\alpha)^4\biggr],
\end{equation}
where $x = ZR_0/a_0$ and $\gamma_E=0.577\ldots$ is Euler's constant.
Results of $Q_0$ are tabulated in Table \ref{tab:tab1}.

\section{\label{sec:derivation}Derivation of radiative correction}

\begin{figure}[H]
\centerline{\includegraphics[scale=0.6]{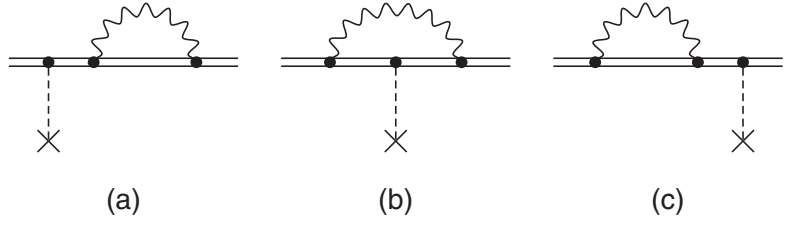}}
\caption{\label{fig:fig1}
Feynman diagrams for the self-energy corrections to parity nonconservation.
The dashed line terminated with a cross indicates an interaction with the
nucleus through the exchange of a Z boson.}
\end{figure}

A principal advantage of treating the degenerate case, where the states
involved in the matrix elements have the same energy, is the simplicity of
the formalism. In the more general case, when the energies are different,
the radiative correction to the weak interaction matrix element has to
involve the laser field photon that drives the transition, otherwise one
would not be dealing with a gauge-invariant amplitude. In the degenerate
case we can restrict our attention to the gauge-invariant subset of diagrams
shown in Fig.~\ref{fig:fig1}, which involve the vertex (Fig.~\ref{fig:fig1}b)
and wave function (Figs.~\ref{fig:fig1}a, \ref{fig:fig1}c) corrections.
While the treatment of these diagrams is straightforward for scattering
processes, more care is required when bound states are involved.
As mentioned in the introduction, the similar problem of radiative
correction to hyperfine splitting has already been treated in the
literature \cite{hfsallZ,BCS}, but in the present case the initial and
final states are different, and the formalism requires some modifications.

The bound state wave functions $\psi_v$ and $\psi_w$ are solutions of
the Dirac equation in the field of a point nucleus. Therefore they can be
interpreted as residues at poles of Dirac-Coulomb propagators as a function
of energy $E\equiv p^0$,
\begin{equation}
S_F(\vec r\,',\vec r,E) = \Big\langle \vec r\,' \Big|
\frac{1}{\not\!p -m-\gamma^0\,V} \Big| \vec r \Big\rangle
\approx \frac{ \psi(\vec r\,') \bar\psi(\vec r)}{E-E_\psi}.
\end{equation}
When radiative corrections are involved, the Dirac-Coulomb propagator is
corrected by the electron self-interaction $\Sigma$
\begin{equation}
\Big\langle \vec r\,' \Big| \frac{1}{\not\! p -m-\gamma^0\,V-\Sigma(E)}
\Big| \vec r \Big\rangle.
\end{equation}
The new position of the pole and corresponding residues are
\begin{eqnarray}
E_\psi^{(1)} ~&=& E_\psi +\langle\bar\psi|\Sigma(E_\psi)|\psi\rangle \\
|\psi\rangle^{(1)} &=& |\psi\rangle +
S'_F(E_\psi)\,\Sigma(E_\psi)|\psi\rangle
+\frac{|\psi\rangle}{2}\,\frac{\partial}{\partial E}\biggr|_{E=E_\psi}
\!\langle\bar\psi|\Sigma(E)|\psi\rangle,
\label{eq:wavefnt}
\end{eqnarray}
where by $S'_F$ one denotes a reduced Coulomb-Dirac propagator, namely the
propagator with the $\psi$-state excluded. With the help of the above
equations, we now present the one-loop radiative corrections to $Q_0$.
They consist of the vertex correction $Q_V$, the left and right wave
function corrections $Q_{SL}+Q_{SR}$, which include as well the
derivative terms, associated with the last term in Eq.~(\ref{eq:wavefnt}).
In the Feynman gauge they are ($\epsilon\equiv E_v=E_w$)
\begin{equation}
Q_V = -4 \pi i \alpha\int\!{d^n k \over (2\pi)^n}\, {1\over k^2+ i \delta}\,
\langle w|e^{i\,\vec k\cdot\vec r}\gamma_{\mu} S_F(\epsilon - k_0)
\gamma_0 \gamma_5 \rho_N S_F(\epsilon - k_0)\gamma^{\mu}
e^{-i\,\vec k\cdot\vec r} |v\rangle,
\label{eq:qv}
\end{equation}
\begin{eqnarray}
Q_{SL} &=& -4 \pi i \alpha\int\!{d^n k \over (2\pi)^n}\,
{1\over k^2+ i \delta} \, \langle w| \gamma_0 \gamma_5 \rho_N
S'_F(\epsilon)e^{i\,\vec k\cdot\vec r}\gamma_{\mu} S_F(\epsilon - k_0)
\gamma^{\mu}e^{-i\,\vec k\cdot\vec r}|v\rangle
\nonumber \\ &&
+2 \pi i \alpha\,Q_0\int\!{d^n k \over (2\pi)^n}\, {1\over k^2+ i \delta}\,
\langle w|e^{i\,\vec k\cdot\vec r}\gamma_{\mu} S_F(\epsilon - k_0)\,
\gamma_0\, S_F(\epsilon - k_0)\gamma^{\mu}e^{-i\,\vec k\cdot\vec r}|w\rangle,
\label{eq:qsl}
\end{eqnarray}
\begin{eqnarray}
Q_{SR} &=&-4 \pi i \alpha\int\!{d^n k \over (2\pi)^n}\,
{1\over k^2+ i \delta}\, \langle w|e^{i\,\vec k\cdot\vec r}\gamma_{\mu}
S_F(\epsilon - k_0)\gamma^{\mu}e^{-i\,\vec k\cdot\vec r} S'_F(\epsilon)
\gamma_0 \gamma_5 \rho_N|v\rangle
\nonumber \\ &&
+2 \pi i \alpha\,Q_0\int\!{d^n k \over (2\pi)^n}\, {1\over k^2+ i \delta}\,
\langle v|e^{i\,\vec k\cdot\vec r}\gamma_{\mu} S_F(\epsilon - k_0)\,\gamma_0\,
S_F(\epsilon - k_0)\gamma^{\mu}e^{-i\,\vec k\cdot\vec r}|v\rangle.
\label{eq:qsr}
\end{eqnarray}
There is still an ambiguity in the above formulas, related to the fact that
at least one of the states is unstable with respect to radiative decay.
This means that, for example, derivative terms, which have the interpretation
of bound state wave function renormalization, acquire a small imaginary part.
We think that this imaginary term may have a small effect on the weak matrix
element. Nevertheless, in our treatment we completely ignore this imaginary
part for simplicity. To include it properly would require a more detailed
treatment of the excitation and decay process. Before the numerical
integration, we present in the next section the analytic calculation of
the first two terms in the $Z\alpha$ expansion.

\section{\label{expansion}{\boldmath $Z\alpha$} expansion}

In the $Z\alpha$ expansion one performs a simplification, similar to
that used for the Lamb shift, which leads to an exact expression for
the expansion terms. Specifically, the first two terms are given by the
on-mass-shell scattering amplitude, which because it involves the weak
charge of the nucleus, is dominated by the large momentum region,
with characteristic momenta of the order of the electron mass, and to
smaller extent of the order of the inverse of nuclear size. The small
momentum region contributes at order $O(Z\alpha)^2$ and will be included
in the numerical treatment. We aim here to confirm the previously obtained
result \cite{Milstein} shown in Eq.~\ref{eq:perturb}, which will be used
later to test the numerical accuracy of the nonperturbative treatment.
In this section we do not pull out a factor $\alpha/\pi$ from $R(Z\alpha)$.

The relative correction to order ${\alpha}$ is determined by considering
the radiative correction  to the $\gamma^\mu\gamma^5$ vertex,
\begin{equation}
\Gamma^{\mu}(p_{2},p_{1})=\frac{{\alpha}}{4\pi}\int{\frac{d^4 q}{{i}{\pi}^2}}
\frac{N^{\mu}(p_{2},p_{1})}{[(q-k)^{2} -m^{2} + i {\epsilon}][q^{2}-m^{2}+
i{\epsilon}][(q-p_{2})^{2}-{\lambda}^{2}+i{\epsilon}]},
\end{equation}
where
\begin{equation}
N^{\mu}(p_{2},p_{1})={\gamma}^{\alpha}(\not\!q  - \not\! k  + m)
{\gamma}^{\mu}{\gamma}^{5}(\not\!q  +m){\gamma}_{\alpha},
\end{equation}
and $k=p_{2}-p_{1}$.
The most general form of $ \Gamma^{\mu} $ in momentum space is
\begin{equation}
\Gamma^{\mu}(p_{2},p_{1})= F_{1}(k^{2})\gamma^{\mu}
{\gamma^{5}}+F_{2}(k^{2})\frac{k^{\mu}}{m} {\gamma}^{5}.
\end{equation}
The form factors $ F_{1}(k^{2})$ and $F_{2}(k^{2})$ are calculated
following the same steps as in the case of the electromagnetic vertex.
Introducing Feynman parameters and taking into account the mass-shell
condition, one obtains in the limit of zero momentum transfer
\begin{eqnarray}
F_{1}(0)&=&-\frac{\alpha}{2\pi} \\  
F_{2}(0)&=&~~\frac{7 \alpha}{12 \pi}.
\end{eqnarray}
For a static nucleus, only $F_{1}(0)$ contributes to the relative
correction to first order in ${\alpha}$. The relative correction
to the PNC amplitude is
\begin{equation}
R(Z\alpha) =\frac{\bar{u}(p,\sigma){\Gamma}^{0}u(p,\sigma)}
{\bar{u}(p,\sigma){\gamma^{0}\gamma^{5}}{u}(p,\sigma)},
\end{equation}
which can be transformed into
\begin{equation}
R(Z\alpha)= \frac{{\rm Tr}\Big[{\Gamma}^{0}\frac{1}{4m}{(\not\!p+m)}
(1\,+\!\not\!a\gamma^5)\Big]} {{\rm Tr}\Big[\gamma^{0}{\gamma}^{5}
\frac{1}{4m}(\not\!p+m) (1\,+\!\not\!a\gamma^5)\Big]},
\end{equation}
where $a^{\mu}=(a_{0},\vec{a})$ with $a\cdot p=0$ is the polarization
four vector of the electron. We then recover the well-known
\cite{Marciano-Sirlin} lowest-order correction
\begin{equation}
R(Z\alpha)=-\frac{\alpha}{2 \pi}.
\end{equation}

\begin{figure}[H]
\centerline{\includegraphics[scale=0.9]{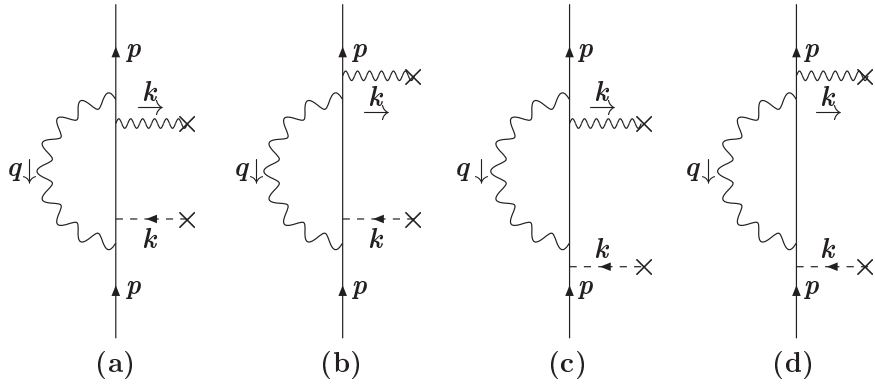}}
\caption{\label{fig:fig2}
Feynman diagrams for the leading binding corrections to the forward
scattering amplitude. The dashed and wavy lines terminated with a cross
represent PNC and Coulomb interactions with the nucleus, respectively.}
\end{figure}

The leading binding correction can be derived from the forward scattering
amplitude, which involves an additional Coulomb exchange. It consists of
the 4 diagrams presented in Figs.~\ref{fig:fig2}a - \ref{fig:fig2}d,
which we evaluate using Yennie gauge. This gauge has the useful property
that each diagram is infrared finite as the photon mass $\lambda$ is taken
to 0. The contribution from Fig.~\ref{fig:fig2}a to the ratio $R(Z\alpha)$
can be written as
\begin{equation}
R_{1}=-\frac{Z\,{\alpha}^{2}}{a_{0}}\int\frac{{d^{3}k}}{(2 \pi)^{3}}
\frac{1}{k^{2}}\int\frac{d^{4}q}{{\pi}^{2} i}\frac{N_{1}(q,k)}
{[q^{2}+i\epsilon]^{2}[(p+q)^{2}-m^{2}+ i\epsilon]^{2}[(p+k+q)^{2}-m^{2}
+ i \epsilon]},
\end{equation}
where
\begin{eqnarray}
\label{eq:Num}
N_{1}(q,k) &=& (g_{\mu \nu}q^{2}+2 q_{\mu}{q_{\nu}})
{\rm Tr}\Big[\gamma^{\mu}(\not\!p\,+\!\not\!q+m){\gamma}^{0}
(\not\!p\,+\!\not\!q\,+\!\not\!k+m){\gamma^{0}{\gamma}^{5}}
\nonumber \\ && \times
(\not\!p\,+\!\not\!q+m){\gamma}^{\nu}\frac{1}{4m}{(\not\!p+m )}
{(1\,+\!\not\!a {\gamma}^{5})}\Big].
\end{eqnarray}
For Figs.~\ref{fig:fig2}b and \ref{fig:fig2}c, we have
\begin{eqnarray}
R_{i}&=&-\frac{Z\,{\alpha}^{2}}{a_{0}}\int\frac{{d^{3}k}}{(2 \pi)^{3}}
\frac{1}{k^{2}}\int\frac{d^{4}q}{{\pi}^{2} i}\frac{1}{[q^{2}+i\epsilon]^{2}}
\nonumber \\ && \times
\frac{N_{i}(q,k)}{[(p+q)^{2}-m^{2}+ i\epsilon][(p+k+q)^{2}-m^{2}+i\epsilon]
[(p+k)^{2}-m^{2}+i{\epsilon}]},
\end{eqnarray}
where
\begin{eqnarray}
N_{2}(q,k)&=&(g_{\mu \nu}q^{2}+2 q_{\mu}{q_{\nu}})
{\rm Tr}\Big[{\gamma}^{0}(\not\!p\,+\!\not\!k + m){\gamma}^{\mu}
(\not\!p\,+\!\not\!k\,+\!\not\!q+m){\gamma}^{0}\gamma^{5}
\nonumber \\ &&
\times(\not\!p\,+\!\not\!q +m){\gamma}^{\nu}\frac{1}{4m}{(\not\!p+m )}
{(1\,+\!\not\!a{\gamma}^{5})}\Big],
\end{eqnarray}
\begin{eqnarray}
N_{3}(q,k)&=&(g_{\mu \nu}q^{2}+2 q_{\mu}{q_{\nu}})
{\rm Tr}\Big[{\gamma}^{\mu}(\not\!p\,+\!\not\!q +m){\gamma}^{0}
(\not\!p \,+\! \not\!k\,+\!\not\!q +m){\gamma}^{\nu}
\nonumber \\ &&
\times(\not\!p \,+\! \not\! k+m)\gamma^{0}\gamma^{5}
{\frac{1}{4m}}(\not\!p+m ){(1\,+\! \not\!a {\gamma}^{5})}\Big].
\end{eqnarray}
Finally, for Fig.~\ref{fig:fig2}d, one has
\begin{equation}
R_{4}=-\frac{Z\,{\alpha}^{2}}{a_{0}}\int\frac{{d^{3}k}}{(2 \pi)^{3}}
\frac{1}{k^{2}}\int\frac{d^{4}q}{{\pi}^{2} i}\frac{N_{4}(q,k)}
{[q^{2}+i\epsilon]^{2}[(p+k)^{2}-m^{2}+ i\epsilon]^{2}
[(p+k+q)^{2}-m^{2}+ i \epsilon]}
\end{equation}
\begin{eqnarray}
N_{4}(q,k)&=&(g_{\mu \nu}q^{2}+2 q_{\mu}{q_{\nu}})
{\rm Tr}\Big[{\gamma}^{0}(\not\!p\,+\!\not\!k+m){\gamma}^{\mu}
(\not\!p\,+\!\not\!q+m){\gamma}^{\nu}(\not\!p\,+\!\not\!k+m){\gamma}^{0}
\nonumber \\ &&
\times{\gamma}^{5}{\frac{1}{4m}}(\not\!p+m )
{(1\,+\! \not\!a {\gamma}^{5})}\Big],
\end{eqnarray}
where $k=(0, \vec{k})$. Each contribution from Figs.~\ref{fig:fig2}a,
\ref{fig:fig2}b, \ref{fig:fig2}c and \ref{fig:fig2}d is written as
\begin{equation}
\label{eq:M}
-\frac{Z\,{\alpha}^{2}}{a_{0}}\!\int\!\frac{d^{3}k}{{(2 \pi)}^{3}}
\frac{F_{i}(k^2)}{k^{2}}.
\end{equation}
The calculations are considerably simplified if one determines only
the imaginary part of the functions $F_{i}(k^{2})$. These are analytic
functions with a branch cut for $ k^{2}> 0 $. The real part of
$F_{i}(k^{2})$ is then obtained by means of Cauchy's theorem,
\begin{equation}
F(k^{2})=\frac{1}{2 \pi i} \int d M^{2}\frac{F(M^{2}+ i 0)- F(M^{2}-i 0)}
{M^2-k^{2}}=\frac{1}{\pi}\int d M^{2}{\frac{{\Im[F(M^{2})]}}{M^2-k^{2}}},
\end{equation}
where $k^{2}<0$. Substituting this expression into Eq.~(\ref{eq:M}) and
integrating over $k$ yields
\begin{equation}
{R_{i}}=\frac{Z\,{\alpha}^{2}}{2 {\pi}^{2} a_{0}}{\int}^{\infty}_{0}{dM}
\Im[F_{i}(M^{2})].
\label{eq:N}
\end{equation}
In order to calculate $\Im[F_{i}]$, a procedure in Mathematica is written
which facilitates the evaluation of the trace in Eq.~(\ref{eq:Num}) and
the the integrals in Eq.~(\ref{eq:N}).  Each contribution is doubled due
to the permutation of photon and boson lines. Setting $m=1$ and picking
the terms linear in ${\vec{p}}$, we obtain
\begin{eqnarray}
\sum_{i=1}^{4}\Im[F_{i}(M^2)] = 2 a_{0} \pi \biggl\{\frac{7}{3}
-\frac{32}{3 M^2}+\frac{2}{3 (1+M^2)^2}-\frac{1}{1+M^2}+
2\,\biggl(\frac{16}{3 M^3}-\frac{1}{M}-\frac{M}{6}\biggr)
\nonumber \\ 
\times\biggl[\arctan(M)-\arccos\Bigl(\frac{2}{M}\Bigr)\theta(M-2)\biggr]
-2\,\biggl(1-\frac{10}{3 M^2}\biggr)\sqrt{1-\frac{4}{M^2}}\,
\theta(M-2)\biggr\},
\end{eqnarray}
where $\theta$ is the step function with $\theta(x)=0$ for $x<0$ and
$\theta(x)=1$ for $x>0$. The above expression can be analytically
integrated. Hence we have
\begin{equation}
R(Z\alpha)
\;=\; -\frac{\alpha}{2 \pi} +  \sum_{i=1}^{4} R_{i}
\;=\; -\frac{\alpha}{2 \pi} + \frac{Z\alpha^2}{2{\pi}^{2}a_{0}}
        \int_{0}^{\infty}\!dM\sum_{i=1}^{4}\Im[F_{i}]
\;=\; -\frac{\alpha}{2 \pi} - \biggl(\frac{7}{12}+
2 {\rm ln} 2 \biggr)Z\alpha^2 ,
\end{equation}
in agreement with Ref.~\cite{Milstein}. We now turn to the numerical
calculation.

\section{\label{sec:numerical}Numerical approach}

In order to make contact with the notation used in Ref.~\cite{BCS},
we note that the two terms $Q_{SL}$ and $Q_{SR}$ in Eqs.~(\ref{eq:qsl})
and (\ref{eq:qsr}) are associated with what are called ``side-left'' (SL)
and ``side-right'' (SR) diagrams in that work, which notation we will
follow in this section. In addition, the SL and SR diagrams have
contributions called ``derivative terms''.  We will refer in this section
to the Gell-Mann Low formalism used in Ref. \cite{BCS} in a rederivation
of Eqs. (\ref{eq:qsl}) and (\ref{eq:qsr}): the adiabatic damping factor
$\epsilon$ used in that formalism can be distinguished from the
factor used in dimensional regularization, $n = 4 - \epsilon$, by context.
In the numerical evaluation,
each diagram breaks into several pieces, which we define as
\begin{equation}
Q = \sum_{i=1}^3 Q_{Vi} + \sum_{i=1}^4 Q_{SLi} + \sum_{i=1}^4 Q_{SRi}.
\end{equation}
We now treat the vertex and side diagrams in turn.

\subsection{Vertex diagram}

The vertex diagram $Q_V$, shown in Fig.~\ref{fig:fig1}b,
was given in Eq.~(\ref{eq:qv}).
The ultraviolet divergent part of the diagram can be isolated by replacing
$S_F$ with $S_0$, where $S_0$ is a free propagator. If this replacement is
made, we get the contribution $Q_{V1}$ which is most conveniently evaluated
in momentum space
\begin{equation}
Q_{V1} = -4 \pi i \alpha \int\! \frac{d^3 p_2}{(2\pi)^3}\,
\frac{d^3 p_1}{(2\pi)^3} \int\! {d^n k \over (2\pi)^n} \,{1 \over k^2 + i \delta}\,
\bar{\psi}_w(\vec p_2) \gamma_{\mu} { 1 \over \not\!p_2 \,-\! \not\!k -m}
V(q) { 1 \over \not\!p_1 \,-\! \not\!k -m} \gamma^{\mu} \psi_v(\vec p_1).
\end{equation}
After Feynman parameterization the $d^n k$ integration can be carried out
with the result
\begin{eqnarray}
Q_{V1} & = & { \alpha \over 2 \pi} \biggl( {C \over \epsilon} -1\biggr)
Q_0 - {\alpha \over 2 \pi} \int_0^1 \!\rho d\rho \int_0^1 \!dx
\!\int\! \frac{d^3 p_2}{(2\,\pi)^3}\, \frac{d^3 p_1}{(2\,\pi)^3} \,
\bar{\psi}_w(\vec p_2) V(q)  \psi_v(\vec p_1)\, {\rm ln}{\Delta_V \over m^2}
\nonumber \\ &&
- \,{\alpha \over 4 \pi} \int_0^1 \!\rho d\rho \int_0^1 \!dx
\!\int\! \frac{d^3 p_2}{(2\,\pi)^3}\, \frac{d^3 p_1}{(2\,\pi)^3}
\nonumber \\ && \times
\Big[\bar{\psi}_w(\vec p_2) \gamma_{\mu} (\not\!p_2 - \not\!Q +m) V(q)
( \not\!p_1 - \not\!Q +m) \gamma^{\mu} \psi_v(\vec p_1) \Big]
{1 \over \Delta_V}.
\label{eq:qv1}
\end{eqnarray}
Here
\begin{eqnarray}
C~ &=& (4 \pi)^{\epsilon /2} \Gamma(1+\epsilon/2), \nonumber \\
Q_{\mu} &=& \rho\, x\, {p_1}_{\mu} + \rho (1-x) {p_2}_{\mu}, \nonumber \\
\Delta_V &=& \rho x ( m^2 - p_1^2) + \rho(1-x) (m^2 -p_2^2) + Q^2,
\nonumber \\
q~ &=& | \vec p_2 - \vec p_1|, \nonumber
\end{eqnarray}
and the Fourier transform of the weak Hamiltonian in the case of a uniform
charge distribution is
\begin{equation}
V(q) = {3 \over 8 \pi^3 (qR_0)^3} \Big[ {\rm sin}(qR_0) - qR_0
{\rm cos}(qR_0)\Big] \gamma_0 \gamma_5.
\end{equation}
The first two terms in the right-hand-side of Eq.~(\ref{eq:qv1})
are divergent and will be held for later cancellation with the
``derivative terms'' from the SL and SR calculation. The remaining
finite parts of $Q_{V1}$ are tabulated in the second column of
Table \ref{tab:tab2}.

The difference of $Q_{V}$ and $Q_{V1}$ is ultraviolet finite, and is
evaluated in coordinate space. The $k_0$ integral is treated by carrying
out a Wick rotation, $k_0 \rightarrow i \omega$, which leads to
\begin{eqnarray}
Q_{V2} &=& -8 \pi \alpha \,\Re \!\int\! d^3 x \!\int\! d^3 y \!\int\! d^3 z
\int_0^{\infty} {d \omega  \over 2\pi}
\int\!{d^3 k \over (2\pi)^3}\,
{e^{i \vec k \cdot(\vec x - \vec z)} \over \omega^2 + \vec k^2 }
\nonumber \\ && \times
\Bigr[\bar{\psi}_w(\vec x) \gamma_{\mu}
S_F(\vec x, \vec y, \epsilon_w - i \omega) \gamma_0 \gamma_5 \rho_N(\vec y)
S_F(\vec y, \vec z; \epsilon_v - i \omega) \gamma^{\mu} \psi_v(\vec z)
\nonumber \\ &&
- \, \bar{\psi}_w(\vec x) \gamma_{\mu}
S_0(\vec x, \vec y, \epsilon_w - i \omega) \gamma_0 \gamma_5 \rho_N(\vec y)
S_0(\vec y, \vec z; \epsilon_v - i \omega) \gamma^{\mu} \psi_v(\vec z)\Bigr].
\end{eqnarray}

A singularity associated with the parts of the bound propagators which
include $w$ or $v$ is regularized by evaluating the expression with
$\epsilon_{w,v} \rightarrow \epsilon_{w,v}(1-\Delta)$ in the electron
propagators. The result behaves as ln$\Delta$. While it is possible to
explicitly cancel this dependence with similar terms from the side
diagrams, we choose here to simply work with a specific, small value of
$\Delta = 10^{-5}$. We note that different choices of $\Delta$ will lead
to slightly different results for $Q_{V2}$, but when combined with the
side diagrams discussed below, the sums are essentially the same as long
as the values of $\Delta$ are reasonably small. Results for $Q_{V2}$ with
$\Delta = 10^{-5}$ are given in the third column of Table \ref{tab:tab2}.

The Wick rotation mentioned above passes bound state poles which must be
accounted for. They are treated by rewriting $Q_V$ by treating the
propagators as a spectral representation, carrying out the $d^3 k$
integration analytically, and defining
\begin{equation}
g_{ijkl}(E) \equiv \alpha \int \! d^3 x \, d^3 y \,
{e^{i \sqrt{E^2+i \delta}\,|\vec x - \vec y\,|} \over |\vec x - \vec y\,|}
\, \bar{\psi}_i (\vec x) \gamma_{\mu} \psi_k(\vec x)
\, \bar{\psi}_j (\vec y) \gamma^{\mu} \psi_l(\vec y),
\end{equation}
which allows us to write
\begin{equation}
Q_V = i \int { dk_0 \over 2 \pi} \sum_{mn} { g_{wnmv}(k_0) Q_{mn} \over
[\epsilon_w( 1 - \Delta) - k_0 - \epsilon_m(1-i\delta)]
[\epsilon_v( 1 - \Delta) - k_0 - \epsilon_n(1-i\delta)]}.
\end{equation}
The choice we have made in regularizing leads to only the ground state
$1s_{1/2}$, denoted as $a$, being encircled when
$k_0 \rightarrow i \omega$, so
\begin{equation}
Q_{V3} =  \sum_{an} { g_{wnav}(\epsilon_w - \epsilon_a) Q_{an}
\over \epsilon_a - \epsilon_n} + \sum_{ma} { g_{wamv}
(\epsilon_v-\epsilon_a) Q_{ma} \over \epsilon_a - \epsilon_m}.
\end{equation}
The sum over $a$ ranges only over the two magnetic quantum numbers
of the state. This contribution is tabulated in the fourth column of
Table \ref{tab:tab2}.  The part of the summation in which the denominator
would vanish corresponds to a double pole, but does not contribute because
$Q_{aa}$ vanishes. However, it should be noted that double poles will
in general contribute, and in fact would be present in the present
calculation were we to use a negative value of $\Delta$ which would
introduce additional pole terms from the $2s_{1/2}$ and $2p_{1/2}$ states.

\subsection{Side diagrams}

It is convenient for the discussion of the side diagrams to introduce
the matrix element of the self-energy operator between two arbitrary
states $m$ and $n$,
\begin{equation}
\Sigma_{mn}(E) = - i e^2  \!\int\! d^3 x d^3 y \!\int\!{d^n k \over
(2 \pi)^n}\, {e^{i \vec k \cdot (\vec x - \vec y)} \over k^2 + i \delta}\,
\bar{\psi}_m(\vec x) \gamma_{\mu} S_F( \vec x, \vec y; E - k_0)
\gamma^{\mu} \psi_n(\vec y).
\end{equation}
A self-mass counterterm is understood to be included in the above.
The self-energy of a valence state is then $\Sigma_{vv}(\epsilon_v)$,
and can be evaluated as described in Ref.~\cite{CJS}.

Using a spectral decomposition of the intermediate propagator, the
$S$-matrix for SL is
\begin{equation}
Q_{SL} = -i \lambda^3 \sum_m\int{d E_1 \over 2 \pi}\int{dE_2 \over 2 \pi}
{Q_{wm} \Sigma_{mv}(E_2)
\over E_1-\epsilon_m(1-i\delta)}\Delta(E_1-\epsilon_w)\Delta(E_1-E_2-k_0)
\Delta(E_2 + k_0 - \epsilon_v),
\end{equation}
and for SR
\begin{equation}
Q_{SR} = -i \lambda^3 \sum_m\int{d E_1 \over 2\pi}\int{dE_2 \over 2\pi}
{\Sigma_{wm}(E_1) Q_{mv}   \over
  E_2 -\epsilon_m(1 - i \delta)}
\Delta(E_2 - \epsilon_v)\Delta(E_1-E_2+k_0)\Delta(E_1+k_0-\epsilon_w),
\end{equation}
with
\begin{equation}
\Delta(E) = { 2 \epsilon \over E^2 + \epsilon^2}.
\end{equation}
Here $\lambda$ is a factor associated with the Gell-Mann-Low formalism
\cite{GML} that is to be differentiated and set to unity: in addition,
a factor $i \epsilon/2$ must be multiplied into the $S$-matrix to obtain
the off-diagonal energy. If the restriction is made that $m \neq w,v$,
it is straightforward to show that two ``perturbed orbital'' (PO)
contributions to the matrix element result which are given by
\begin{equation}
Q_{SL1} =  \sum_{m \neq v} { Q_{wm} \Sigma_{mv}(\epsilon_v) \over
\epsilon_w - \epsilon_m} \equiv \Sigma_{\tilde{v} v}(\epsilon_v)
\end{equation}
and
\begin{equation}
Q_{SR1} = \sum_{m \neq w} { \Sigma_{wm}(\epsilon_w) Q_{mv}  \over
\epsilon_v - \epsilon_m} \equiv \Sigma_{w \tilde{w}}(\epsilon_w).
\end{equation}
This is equivalent to the forms given for $Q_{SL}$ and $Q_{SR}$ in
Eqs.~(\ref{eq:qsl}) and (\ref{eq:qsr}). We note that it is not necessary
to explicitly make the restrictions $m \neq w$ in $Q_{SL1}$ and $m \neq v$
in $Q_{SR1}$ because $Q_{ww} = Q_{vv} = 0$. The PO terminology arises from
the fact that the $m$ summation can be carried out before evaluating the
self-energy, and one then needs only to do a self-energy calculation with
one of the external wavefunctions replaced with a perturbed orbital.
The PO terms are tabulated in the fifth and ninth columns of
Table \ref{tab:tab2}.

The cases $m=v$ and $m=w$ are more subtle, as they contribute terms of
order $1/ \epsilon$ to the off-diagonal energy. This divergence cancels,
but a finite contribution coming from Taylor expanding $\Sigma(E)$
remains, and contributes ${1 \over 2} Q_0 {\Sigma}'_{vv}(\epsilon_v)+
{1 \over 2} Q_0 {\Sigma}'_{ww}(\epsilon_w)$, or more explicitly
\begin{eqnarray}
Q^{\rm der}_{SL} & = & 2 i \pi \alpha Q_0 \int d^3 y d^3 z d^3 w
\int {d^n k \over (2\pi)^n} {e^{i \vec k \cdot(\vec y - \vec z)}
\over k^2 + i \delta}
\nonumber \\ &&
\times \bar{\psi}_v(\vec y) \gamma_{\mu}
S_F(\vec y, \vec w; \epsilon_v - k_0) \gamma_0
S_F(\vec w, \vec z; \epsilon_v - k_0) \gamma^{\mu} \psi_v(\vec z)
\nonumber \\ & = & -{i \over 2} Q_0 \sum_m \int {d k_0 \over 2 \pi}
{ g_{vmmv}(k_0) \over (\epsilon_v - k_0 - \epsilon_m)^2}
\end{eqnarray}
and
\begin{eqnarray}
Q^{\rm der}_{SR} & = & 2 i \pi \alpha Q_0 \int d^3 y d^3 z d^3 w
\int {d^n k \over (2\pi)^n} {e^{i \vec k \cdot(\vec y - \vec z)}
\over k^2 + i \delta}
\nonumber \\ &&
\times \bar{\psi}_w \gamma_{\mu}
S_F( \vec y, \vec w; \epsilon_w - k_0) \gamma_0
S_F(\vec w, \vec z; \epsilon_w - k_0) \gamma^{\mu} \psi_w(\vec z)
\nonumber \\ & = & -{i \over 2} Q_0 \sum_m \int {d k_0 \over 2 \pi}
{ g_{wmmw}(k_0) \over (\epsilon_w - k_0 - \epsilon_m)^2}.
\end{eqnarray}
These correspond to the derivative term mentioned in the preceding
section. The analysis of this term parallels closely the treatment
of the vertex: first the bound propagators are replaced with free
propagators, which gives
\begin{equation}
Q_{SL2} = 2 \pi i \alpha Q_0
\!\int\! \frac{d^3 p}{(2\,\pi)^3} \!\int\! {d^n k \over (2\pi)^n}\,
\bar{\psi}_v(\vec p) \gamma_{\mu} {1 \over \not\!p \,-\! \not\!k - m}
\gamma_0 {1 \over \not\!p \,-\! \not\!k - m} \gamma^{\mu} \psi_v(\vec p)
\end{equation}
and
\begin{equation}
Q_{SR2} = 2 \pi i \alpha Q_0 \!\int\! \frac{d^3 p}{(2\,\pi)^3}
\!\int\! {d^n k \over (2\pi)^n}\, \bar{\psi}_w(\vec p)
\gamma_{\mu} {1 \over \not\!p \,-\! \not\!k - m} \gamma_0
{1 \over \not\!p \,-\! \not\!k - m} \gamma^{\mu} \psi_w(\vec p).
\end{equation}
Feynman parameterizing and carrying out the $d^n k$ integration gives
\begin{eqnarray}
Q_{SL2} &=& -{ \alpha \over 4 \pi} \biggl( {C \over \epsilon} -1\biggr)
Q_0 + {\alpha \over 4 \pi} Q_0 \int_0^1\! \rho d\rho \int\! \frac{d^3 p}
{(2\,\pi)^3}\, \bar{\psi}_v(\vec p) \gamma_0  \psi_v(\vec p)
{\rm ln}{\Delta_S \over m^2}
\nonumber \\ && \hspace{-1.4em}
+ {\alpha \over 8 \pi} Q_0 \int_0^1\! \rho d\rho \!\int\!
\frac{ d^3 p}{(2\,\pi)^3} \Big\{\bar{\psi}_v(\vec p) \gamma_{\mu}
\big[\!\not\!p(1-\rho) +m\big] \gamma_0
\big[\!\not\!p(1-\rho) +m\big] \gamma^{\mu} \psi_v(\vec p)\Big\}
{1 \over \Delta_S}.
\end{eqnarray}
and
\begin{eqnarray}
Q_{SR2} &=& -{ \alpha \over 4 \pi} \biggl( {C \over \epsilon} -1\biggr)
Q_0 + {\alpha \over 4 \pi} Q_0 \int_0^1\! \rho d\rho \int\! \frac{ d^3 p}
{(2\,\pi)^3}\, \bar{\psi}_w(\vec p) \gamma_0  \psi_w(\vec p)
{\rm ln}{\Delta_S \over m^2}
\nonumber \\ && \hspace{-1.4em}
+ {\alpha \over 8 \pi} Q_0 \int_0^1\! \rho d\rho \!\int\!
\frac{ d^3 p}{(2\,\pi)^3} \Big\{\bar{\psi}_w(\vec p) \gamma_{\mu}
\big[\!\not\!p(1 - \rho) +m\big] \gamma_0
\big[\!\not\!p(1 - \rho) +m\big] \gamma^{\mu} \psi_w(\vec p)\Big\}
{1 \over \Delta_S}.
\end{eqnarray}
where $\Delta_S = \rho^2 p^2 +\rho (m^2 - p^2)$. The first two terms in
the right-hand-side of the above equations for $Q_{SL2}$ and $Q_{SR2}$
are divergent but cancel with the corresponding terms of $Q_{V1}$ in
Eq.~(\ref{eq:qv1}). The remaining, finite terms are presented in the
sixth and tenth columns of Table \ref{tab:tab2}.

The difference of the side diagrams evaluated with bound propagators and
free propagators is again ultraviolet finite, and after Wick rotation one has
\begin{eqnarray}
Q_{SL3} &=& 2 \pi Q_0 \alpha \,\Re \!\int\! d^3 x \!\int\! d^3 y \!\int\!
d^3 z \!\int\! {d \omega  \over 2 \pi} \!\int\!{d^3 k \over (2\pi)^3}\,
{e^{i \vec k \cdot(\vec x - \vec z)} \over \omega^2 + \vec k^2 }
\nonumber \\ && \times
\biggl[\bar{\psi}_v(\vec x) \gamma_{\mu}
S_F(\vec x, \vec y, \epsilon_v - i \omega) \gamma_0
S_F(\vec y, \vec z; \epsilon_v - i \omega) \gamma^{\mu} \psi_v(\vec z)
\nonumber \\ &&
-\, \bar{\psi}_v(\vec x) \gamma_{\mu}
S_0(\vec x, \vec y, \epsilon_v - i \omega) \gamma_0
S_0(\vec y, \vec z; \epsilon_v - i \omega) \gamma^{\mu} \psi_v(\vec z)
\biggr].
\end{eqnarray}
and
\begin{eqnarray}
Q_{SR3} &=& 2 \pi Q_0 \alpha \,\Re \!\int\! d^3 x \!\int\! d^3 y \!\int\!
d^3 z \!\int\! {d \omega  \over 2 \pi} \!\int\!{d^3 k \over (2\pi)^3}\,
{e^{i \vec k \cdot(\vec x - \vec z)} \over \omega^2 + \vec k^2 }
\nonumber \\ && \times
\biggl[\bar{\psi}_w(\vec x) \gamma_{\mu}
S_F(\vec x, \vec y, \epsilon_w - i \omega) \gamma_0
S_F(\vec y, \vec z; \epsilon_w - i \omega) \gamma^{\mu} \psi_w(\vec z)
\nonumber \\ &&
-\, \bar{\psi}_w(\vec x) \gamma_{\mu}
S_0(\vec x, \vec y, \epsilon_w - i \omega) \gamma_0
S_0(\vec y, \vec z; \epsilon_w - i \omega) \gamma^{\mu} \psi_w(\vec z)
\biggr].
\end{eqnarray}
The same regularization of the valence energy,
$\epsilon_v \rightarrow \epsilon_v (1-\Delta)$ used in the vertex is
required, and again we simply use $\Delta = 10^{-5}$ and present the
results in the seventh and eleventh columns of Table \ref{tab:tab2}.

Finally, the Wick rotation passes a double pole when $m=a$, with $a$
being the $1s_{1/2}$ ground state, leading to the derivative terms
\begin{equation}
Q_{SL4}= {1 \over 2} Q_0 \sum_a g'_{vaav}(\epsilon_v - \epsilon_a)
\end{equation}
and
\begin{equation}
Q_{SR4}= {1 \over 2} Q_0 \sum_a g'_{waaw}(\epsilon_w - \epsilon_a)
\end{equation}
which are tabulated in the eighth and twelfth column of Table \ref{tab:tab2}.
This completes the calculation and the sums of vertex and side diagram
contributions give the values of the exact evaluation of the function
$R(Z\alpha)$ which are tabulated in the last column of Table \ref{tab:tab2}.

\section{discussion}

\begin{figure}[H]
\centerline{\includegraphics[scale=0.6]{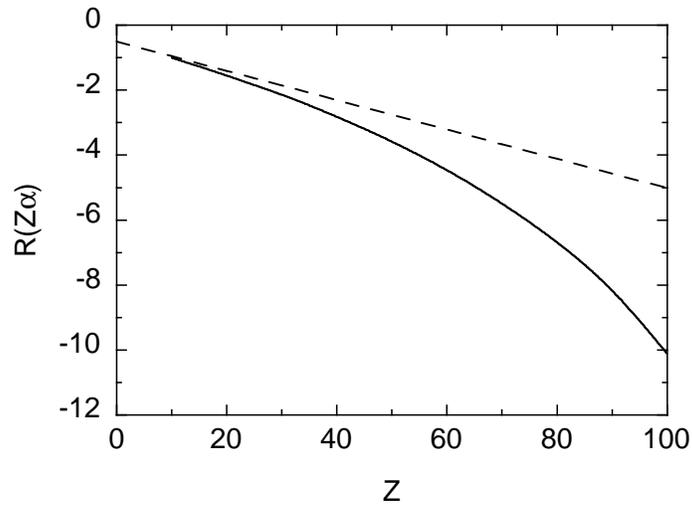}}
\caption{\label{fig:fig3}
Comparison of exact calculation (solid line) with first two terms of
perturbative expansion for $R(Z\alpha)$ (dashed line).}
\end{figure}

A number of numerical issues arise in the calculation that we note here.
In some parts the use of a uniform distribution, with its step function
behavior, caused loss of accuracy. In those cases a fermi distribution was
used: while this leads to small changes in $Q_0$, the effect on $R(Z\alpha)$
is negligible. More serious is the difficulty of controlling numerical
instabilities at low $Z$, which led to our choosing the lowest $Z$ to
be 10. A graph of the numerical value of $R(Z\alpha)$, along with results
from the two leading terms given in Eq.~\ref{eq:perturb}, is shown in
Fig.~\ref{fig:fig3}. The accuracy of the calculation at low $Z$ is
sufficient to allow a fit that determines
\begin{equation}
R(Z\alpha)_{\rm fit}  = - { 1 \over 2} - 0.045(2) Z.
\end{equation}
which is in agreement with the numerical value of Eq.~(\ref{eq:perturb})
\begin{equation}
R(Z\alpha)  = - { 1 \over 2} - 0.045154 Z.
\end{equation}

This agreement provides a check on the rather complex numerical calculation.
The advantage of the numerical approach is of course the fact that it does
not assume $Z \alpha$ to be a small parameter, and thus can be used for
high $Z$. In the particularly interesting case of $Z=55$, we see that
the perturbative formula happens to be -2.983, as compared with the exact
result -4.007.

However, at this point we make no claims about the applicability of the
present calculation to the case of PNC transitions in neutral cesium.
The actual process studied in the experiment that measures PNC \cite{Wieman}
involves a double perturbation, where not only the weak Hamiltonian but
also an external laser photon field act to either first transform the
$6s_{1/2}$ electron into a state with the opposite parity, followed by
an allowed dipole transition to a $7s_{1/2}$ electron, or vice-versa.
To extend our calculation to the actual experiment requires the following
steps.

\begin{figure}[H]
\centerline{ \includegraphics[scale=0.6]{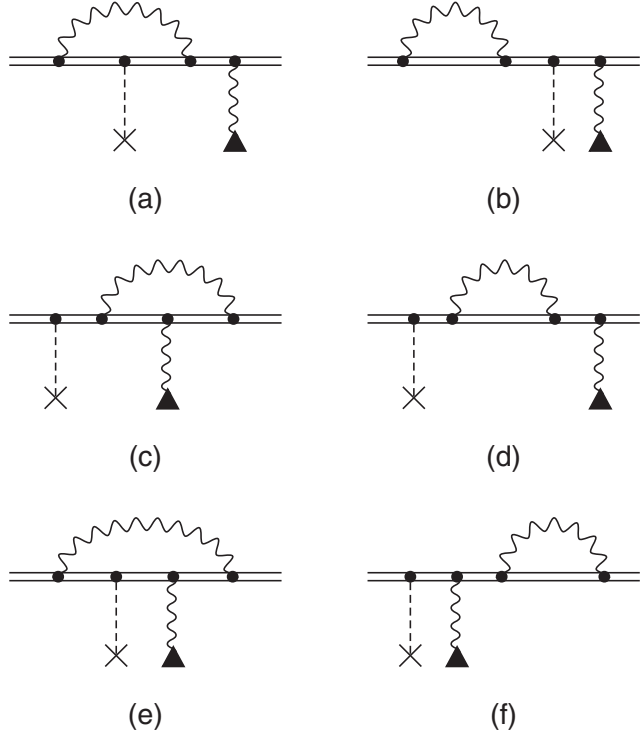}}
\caption{\label{fig:fig4}
Feynman diagrams for the radiative correction to electron excitation
by a laser photon, indicated by the wavy line terminated with a triangle,
in the presence of interaction with the nucleus through exchange of
a Z boson, indicated by the dashed line terminated with a cross.}
\end{figure}

The first step is replacing the Coulomb wave functions used here with
realistic wave functions for neutral cesium. The technology to carry out
radiative corrections in neutral atoms has only recently been put into
place. It is now possible, using a local potential that incorporates
screening, to carry out accurate self-energy \cite{Pyykko,CS} and
radiative correction to hfs \cite{NewCS}  calculations. The second step
is to incorporate the laser photon. This is a more complicated task,
since the set of diagrams shown in Fig.~\ref{fig:fig4} must be evaluated.
We note that while Fig.~\ref{fig:fig4}a corresponds to the vertex
correction considered here, Fig.~\ref{fig:fig4}c corresponds to a
radiative correction to the electromagnetic vertex, and
Fig.~\ref{fig:fig4}e to a new radiative correction specific to
the experiment. We expect that the new contributions will affect
$R(Z\alpha)$ in order $(Z\alpha)^2$, but until they are explicitly
evaluated, their importance for cesium PNC is unknown.

The principal results of this paper are then as follows. Firstly, an
independent scattering calculation of the leading binding correction
in the function $R(Z \alpha)$ has been presented, which confirms the
calculation of Ref.~\cite{Milstein}. Secondly, it has been shown that
numerical methods that allow the evaluation of $R(Z\alpha)$ to all
orders in $Z\alpha$ for the case of gauge-invariant $2s_{1/2}-2p_{1/2}$
transitions in hydrogenlike ions can be applied. The behavior of the
function shows that large binding corrections are present. If these
corrections behave the same in the realistic cesium case, and the
extra corrections of order $(Z \alpha)^2$ mentioned above are small,
the apparent $2 \sigma$ discrepancy noted in Ref.~\cite{vacpol} is
reduced, but before a complete calculation is completed the theoretical
status of PNC in cesium should be regarded as unresolved.

\begin{acknowledgments}
The work of J.S. was supported in part by NSF Grant No.~PHY-0097641.
The work of K.P and A.V was supported by EU Grant No.~HPRI-CT-2001-50034.
The work of K.T.C. was performed under the auspices of the U.S. Department
of Energy by Lawrence Livermore National Laboratory under Contract
No.~W-7405-Eng-48.
\end{acknowledgments}

\begin{table}[H]
\caption{\label{tab:tab1}
Nuclear parameters $c$ and $R_0$ and lowest-order PNC matrix element
$Q_0$: units of fermis for $c$ and $R_0$ and $1/a_0^3$ for $Q_0$.
Square brackets indicate power of 10.}
\begin{ruledtabular}
\begin{tabular}{rrrr}
   \multicolumn{1}{c}{$Z$}
& \multicolumn{1}{c}{$c$}
& \multicolumn{1}{c}{$R_0$}
& \multicolumn{1}{c}{$Q_0$} \\
\colrule
10   &  2.9889 & 3.859 & 1.318[0] \\
15   &  3.2752 & 4.127 & 7.038[0] \\
20   &  3.7188 & 4.487 & 2.388[1] \\
25   &  4.0706 & 4.783 & 6.366[1] \\
30   &  4.4454 & 5.106 & 1.465[2] \\
40   &  4.9115 & 5.516 & 5.988[2] \\
50   &  5.4595 & 6.010 & 2.010[3] \\
55   &  5.6748 & 6.206 & 3.539[3] \\
60   &  5.8270 & 6.345 & 6.136[3] \\
70   &  6.2771 & 6.761 & 1.786[4] \\
80   &  6.6069 & 7.068 & 5.184[4] \\
90   &  6.9264 & 7.368 & 1.542[5] \\
100  &  7.1717 & 7.599 & 4.886[5] \\
\end{tabular}
\end{ruledtabular}
\end{table}

\begin{table*}[H]
\caption{\label{tab:tab2}Breakdown of Contributions to $R(Z\alpha)$.}
\begin{ruledtabular}
\begin{tabular}{crrrrrrrrrrrr}
   \multicolumn{1}{c}{$Z$}
& \multicolumn{1}{c}{$Q_{V1}$}
& \multicolumn{1}{c}{$Q_{V2}$}
& \multicolumn{1}{c}{$Q_{V3}$}
& \multicolumn{1}{c}{$Q_{SL1}$}
& \multicolumn{1}{c}{$Q_{SL2}$}
& \multicolumn{1}{c}{$Q_{SL3}$}
& \multicolumn{1}{c}{$Q_{SL4}$}
& \multicolumn{1}{c}{$Q_{SR1}$}
& \multicolumn{1}{c}{$Q_{SR2}$}
& \multicolumn{1}{c}{$Q_{SR3}$}
& \multicolumn{1}{c}{$Q_{SR4}$}
& \multicolumn{1}{c}{$R(Z\alpha)$} \\
\colrule
  10 & -2.500 &  4.260 &-11.768 & -0.272 & 2.369 & 2.293 & 0.002 & 
-0.045 & 2.269 & 2.388 & 0.006 & -0.998 \\
  15 & -2.009 & -0.253 & -7.853 & -0.415 & 1.972 & 2.689 & 0.003 & 
-0.068 & 1.873 & 2.782 & 0.009 & -1.270 \\
  20 & -1.729 & -2.600 & -5.898 & -0.555 & 1.694 & 2.967 & 0.004 & 
-0.097 & 1.595 & 3.058 & 0.012 & -1.549 \\
  25 & -1.559 & -4.064 & -4.727 & -0.696 & 1.482 & 3.180 & 0.005 & 
-0.128 & 1.384 & 3.269 & 0.014 & -1.840 \\
  30 & -1.458 & -5.067 & -3.948 & -0.839 & 1.311 & 3.350 & 0.006 & 
-0.164 & 1.214 & 3.439 & 0.017 & -2.139 \\
  40 & -1.377 & -6.398 & -2.980 & -1.139 & 1.048 & 3.620 & 0.008 & 
-0.279 & 0.952 & 3.703 & 0.022 & -2.820 \\
  50 & -1.386 & -7.267 & -2.406 & -1.461 & 0.851 & 3.824 & 0.011 & 
-0.434 & 0.755 & 3.904 & 0.026 & -3.583 \\
  55 & -1.411 & -7.609 & -2.201 & -1.637 & 0.768 & 3.911 & 0.012 & 
-0.530 & 0.673 & 3.989 & 0.028 & -4.007 \\
  60 & -1.446 & -7.912 & -2.033 & -1.821 & 0.694 & 3.990 & 0.013 & 
-0.644 & 0.600 & 4.067 & 0.030 & -4.462 \\
  70 & -1.525 & -8.466 & -1.778 & -2.229 & 0.566 & 4.132 & 0.016 & 
-0.922 & 0.472 & 4.207 & 0.033 & -5.494 \\
  80 & -1.617 & -8.922 & -1.602 & -2.707 & 0.458 & 4.253 & 0.019 & 
-1.300 & 0.365 & 4.331 & 0.035 & -6.687 \\
  90 & -1.708 & -9.376 & -1.489 & -3.288 & 0.364 & 4.366 & 0.023 & 
-1.822 & 0.273 & 4.448 & 0.035 & -8.174 \\
100 & -1.798 & -9.831 & -1.433 & -4.020 & 0.282 & 4.457 & 0.028 & 
-2.570 & 0.192 & 4.545 & 0.033 &-10.115 \\
\end{tabular}
\end{ruledtabular}
\end{table*}

\end{document}